\documentclass[aps,prl,twocolumn,superscriptaddress, showpacs]{revtex4}

\usepackage{graphicx}
\usepackage{float}
\usepackage{floatflt}
\DeclareGraphicsExtensions{.jpg,.pdf,.png,.eps}

\begin{document}

\title{Phonon Heat Conduction in Corrugated Silicon Nanowires Below the Casimir Limit}

\author{Christophe Blanc}
\affiliation{Institut N\'EEL, CNRS-UJF, 25 avenue des Martyrs, 38042 Grenoble Cedex 9, France}

\author{Ali Rajabpour}
\affiliation{Department of Mechanical Engineering, Imam Khomeini International University, Qazvin 34149-16818, Iran}

\author{Sebastian Volz}
\affiliation{Laboratoire d'Energ\'etique Mol\'eculaire et Macroscopique, Ecole Centrale Paris, Grande Voie des Vignes, Chatenay Malabry 92295, France}

\author{Thierry Fournier}
\affiliation{Institut N\'EEL, CNRS-UJF, 25 avenue des Martyrs, 38042 Grenoble Cedex 9, France}

\author{Olivier Bourgeois}
\affiliation{Institut N\'EEL, CNRS-UJF, 25 avenue des Martyrs, 38042 Grenoble Cedex 9, France}

\date{\today}


\begin{abstract}

The thermal conductance of straight and corrugated monocrystalline silicon nanowires has been measured between 0.3~K and 5~K. It is demonstrated that the corrugation strongly reduces the thermal transport by reducing the mean free path of the phonons. The experimental averaged mean free path is remarkably smaller than the smaller diameter of the nanowire, evidencing a phonon thermal transport reduced below the Casimir limit. Monte Carlo simulations highlight that this effect can be attributed to significant multiple scattering of ballistic phonons occuring on the corrugated surfaces. This result suggests an original approach to transforming a monocrystalline material into a phonon glass.
\end{abstract}

\maketitle

In the past decade, the thermal and thermodynamic properties of nanostructured materials or low dimensional materials has attracted growing interest \cite{Cahill2003,mingoNat,Lishi,balandin}, especially for applications in thermoelectrics \cite{shakouriTAP,shakouriReview,hochbaum}. Recent experiments and theoretical research have shown the potential for engineering phonons at the nanoscale, for instance, the reduction in phonon thermal conductivity in nanostructured thin films, nanowires  \cite{donadio,PRBHeron,Heron1,Heron2,mingoNat}. Overall, size effects on the heat conduction have been widely studied, demonstrating effects on the phonon mean free path (MFP) \cite{chen2011,cuffe}, on the dispersion relation \cite{PRBMingo2003}, or on the transmission coefficient \cite{volz,donadio}. In general, reducing the phonon thermal transport below the Casimir limit, where the phonon mean free path is limited by boundary scattering, remains an experimental challenge \cite{carrete,wang,Lim,wingert2}. All these recent experiments have involved the measurement of systems with nanoscale sizes, which has motivated the development of specific fine thermal experiments \cite{shi2003,PRLBourg,APLSadat,wingert,sikora}.

In reduced size systems, surface scattering of phonons plays a crucial role \cite{casimir,ziman}. As a result, effects linked to surface roughness have been at the core of numerous studies showi ng a significant impact on the phonon transport especially at low temperatures \cite{holmes,fon,Heron1,Heron2,chen,martin,liu,Oh}. Recent experiments have demonstrated very low thermal conductances of highly rough silicon nanowires \cite{hochbaum,Lim}, far below the amorphous limit. Hence, the possibility of creating a phonon glass type material from a perfect monocrystal seems possible by playing on the geometry of the surface. These nanoengineered surfaces may introduce phonon backscattering or multiple scattering trapping phonons in a lattice of cavities \cite{sawtooth}. In this respect, phononic crystals (a regular modulation of the geometry of a phonon waveguide) have attracted significant attention in the past decade as a promising means to reduce phonon transport \cite{cleland,numa,tang,yu,hopkins}. Despite numerous theoretical works, no measurement has been yet reported demonstrating phonon transport below the Casimir limit in periodic (corrugated) phonon conductor.

Here, we report experimental evidence of a strong reduction in the phonon mean free path (MFP), below the Casimir limit, in silicon corrugated nanowires. The phonon flux is estimated through the measurement of the  thermal conductance of straight and corrugated suspended nanowires. The modulated surfaces induce a severe reduction in the MFP, which becomes significantly smaller than the minimum diameter. Experimental data for the mean free path are compared with predictions based on the Monte Carlo technique, which involves multiple scattering.

\begin{figure}
\begin{center}
 \includegraphics[width=6cm]{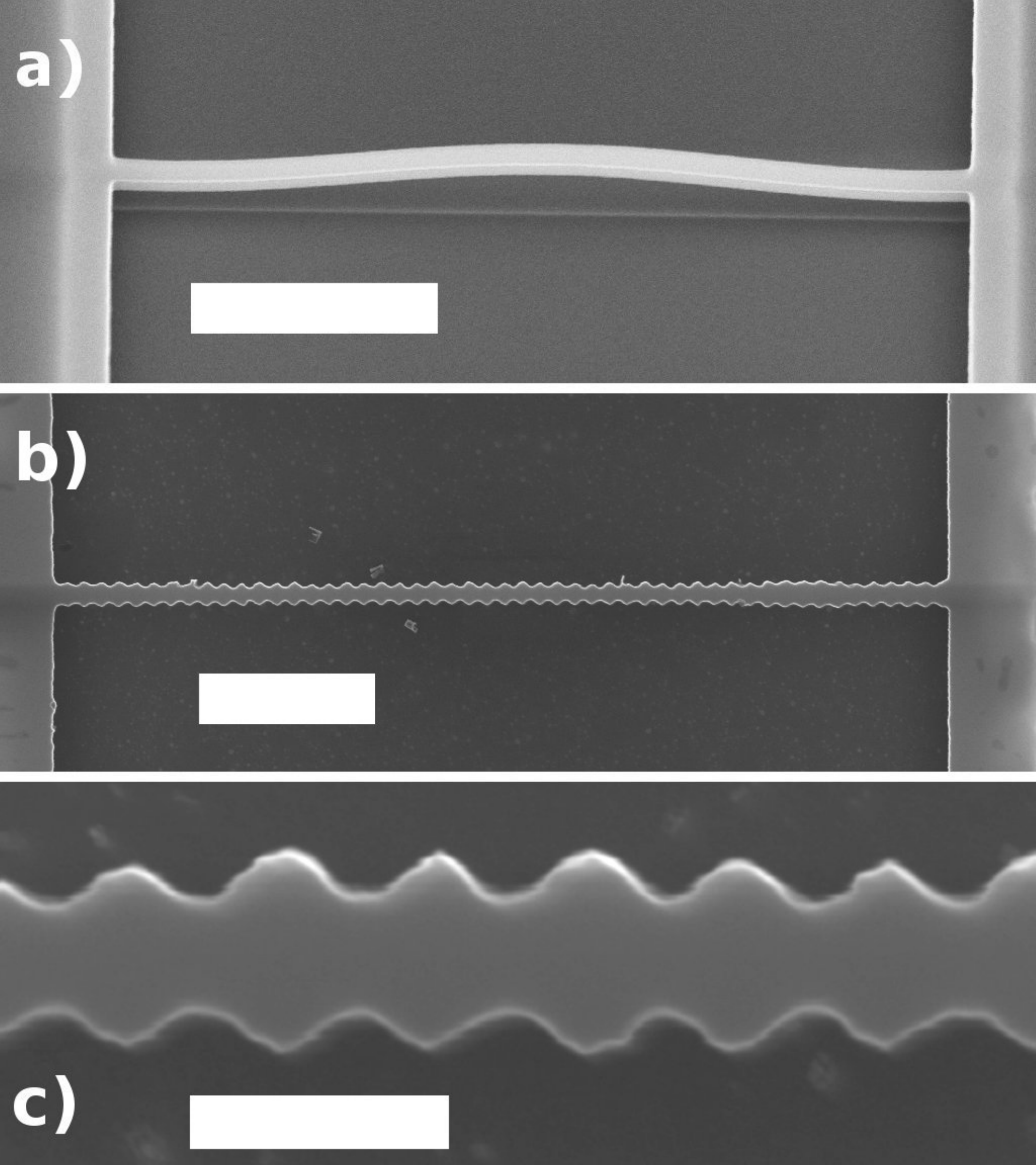}
 \end{center}
 \caption{SEM images of the straight a) and the corrugated b) nanowires; c) corresponds to the top view of the corrugated nanowire. The scale bars correspond to a) 2~$\mu$m, b) 2~$\mu$m and c) 300~nm.} 
 \label{fig1}
\end{figure}

The thermal conductance of suspended monocrystalline silicon nanowires has been measured for temperatures ranging between 0.3~K and 5~K. The Si nanowires (straight and corrugated) are made by high resolution e-beam lithography from intrinsic silicon on insulator. The minimal cross section of the corrugated nanowire is 150~nm$\times$200~nm and the maximal one is 250~nm$\times$200~nm \cite{note1}. The semi-length of the nanowires is either 4.4 $\mu$m or 5 $\mu$m \cite{note3} and the roughness is estimated to be of a few nanometers \cite{Heron1}. Scanning Electron Microscope images of the measured nanowires are shown in Fig.~\ref{fig1}. The periodicity of the corrugation is 200~nm. This length as well as the width is still larger than the dominant phonon wavelength, estimated to be on the order of 65~nm at 1~K, and 6.5~nm at 10K : $\lambda_{dom}=\frac{h \ v_{s}}{4.25 \  k_{B} \ T},$ where \textit{h} is the Planck constant, $v_{s}$ refers to the speed of sound, which was set to 6500~m/s (the mean value of the speed of sound of the transverse and longitudinal acoustic phonons), and $k_{B}$ is the Boltzmann constant \cite{pohl,pokatilova}. between 1~K and 10~K, all the sample dimensions are at least two times bigger than $\lambda_{dom}$ excluding significant effect of phonon confinement on the thermal properties.

The experiments were performed by using the 3$\omega$ method \cite{Cahill,LuRSI,JAPBourg}. This measurement technique has already been explained in detail \cite{Heron1,JAPBourg}. A highly sensitive thermometer in niobium nitride is deposited on top of the intrinsic suspended silicon nanowires \cite{RSIBourg}. An RMS current smaller than $1$~nA is used to create a temperature oscillation in the thermometer resistance (on the order of 600~kOhm at 1~K). This leads to a dissipated power ranging from $10^{-15}$~Watt to $10^{-13}$~Watt; a temperature oscillation amplitude of 3 to 10~mK is thus generated in the temperature range of 0.3~K to 6~K.

The measured thermal conductance values are spread over three orders of magnitude between $10^{-13}$~W/K to $10^{-10}$~W/K. All the measurements were  made at a frequency below 5~Hz, where the nanowires are in the quasistatic thermal regime \cite{JAPBourg}. In this frequency range, the thermal signal is only sensitive to the thermal conductance; the experiment has a resolution estimated to be a few percentage points with a high energy resolution (down to the attoJoule ($10^{-18}$~J)) \cite{note2}. The parasitic thermal path through the thermometer can be estimated by the Wiedemann-Franz law giving a thermal conductance of $4.10^{-14}$~W/K at 1K, much below the thermal conductance of the nanowires.

\begin{figure} [t]
\begin{center}
 \includegraphics[width=8.5cm]{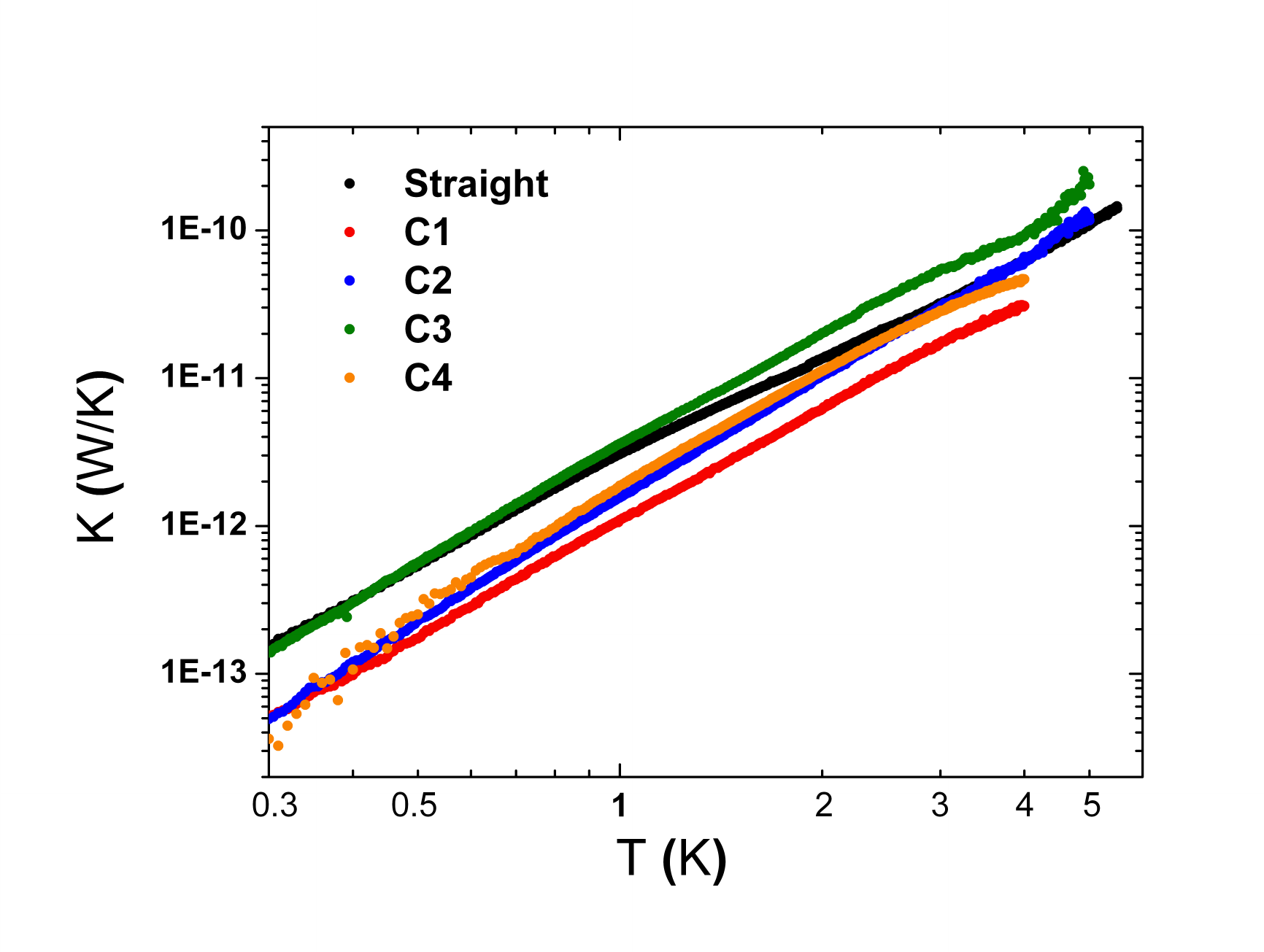}
 \end{center}
 \caption{(colour online) Thermal conductance versus temperature for a straight nanowire and four corrugated nanowires in the log-log scale.} 
 \label{fig2}
\end{figure}

In Fig.~\ref{fig2}, we show the thermal conductance data versus temperature for the corrugated (samples C1 to C4) and the straight nanowire. Table~\ref{tablo} gathers all experimental data and nanowire characteristics. The values obtained for the straight nanowire serve as a reference when analysing the thermal properties of the corrugated nanowires. In order to compare the nanowire thermal conductances to each other, they will be normalized with respect to the one predicted by the Casimir model. In the Casimir theory adapted by Ziman and coworkers \cite{casimir,berman}, the thermal conductance is given by:
\begin{equation}
K_{Cas}=3.2 \times 10^{3} \left( \frac{2\pi ^{2}k_{B}^{4}}{5 \hbar^{3} v_{s}^{3}} \right) ^{(2/3)} \frac{e\times w \Lambda_{Cas}}{L}T^{3}=\beta_{Cas} T^{3},
\label{Caseq}
\end{equation}
$\Lambda_{Cas}=1.12 \sqrt{e\times w}$ is the phonon Casimir MFP for a rectangular shape phonon conductor, $e$ refers to the thickness and $w$ to the average width of the nanowire, $L$ being its length \cite{PRBHeron}. This formula is obtained when the mean free path of the phonons is limited by the cross section of the wire, i.e., by boundary scattering. 

\begin{table*}
\begin{tabular}{|c|c|c|c|c|c|c|c|c|}
  \hline
  Sample & L ($\mu$m)& w$_{min} (nm)$ & w$_{max} (nm)$ & period (nm)& MFP (nm) & Normalized MFP &b & p\\
  \hline
  \hline
  straight & 5 & 200 &200 &0&454 & 1 & 2.27& 0.60 \\
  \hline
  C1&4.4 &130 &250 &200 &57 & 0.13 & 2.52&-0.32\\
  \hline
  C2&5 &140 &245 &200 &90 & 0.2 & 2.74&-0.10\\
  \hline
  C3&5 &160 &260 &200 &178 & 0.39 & 2.54&0.22\\
  \hline
  C4&4.4 &230 &250 & 50 &70 & 0.15 & 2.6&-0.27\\
  \hline
\end{tabular}
\caption{Geometrical parameters and physical data extracted from the thermal conductance measurements of the straight and corrugated nanowires (C$_i$). w$_{min}$ and w$_{max}$ designate the minimum and maximum width of the corrugated nanowires and $L$ their length. The thickness of the silicon wires is equal to 200~nm.}
\label{tablo}
\end{table*}

As the temperature decreases, the dominant phonon wavelength actually grows larger than the roughness height and specular reflections become more probable. To take this trend into account, Ziman proposed replacing the Casimir mean free path by an effective MFP $\Lambda_{eff}$ \cite{ziman,berman}:
\begin{equation}
\Lambda_{eff}=\frac{1+p}{1-p} \Lambda_{Cas},
\label{zimfp}
\end{equation}

\begin{figure}
\begin{center}
 \includegraphics[width=8.5cm]{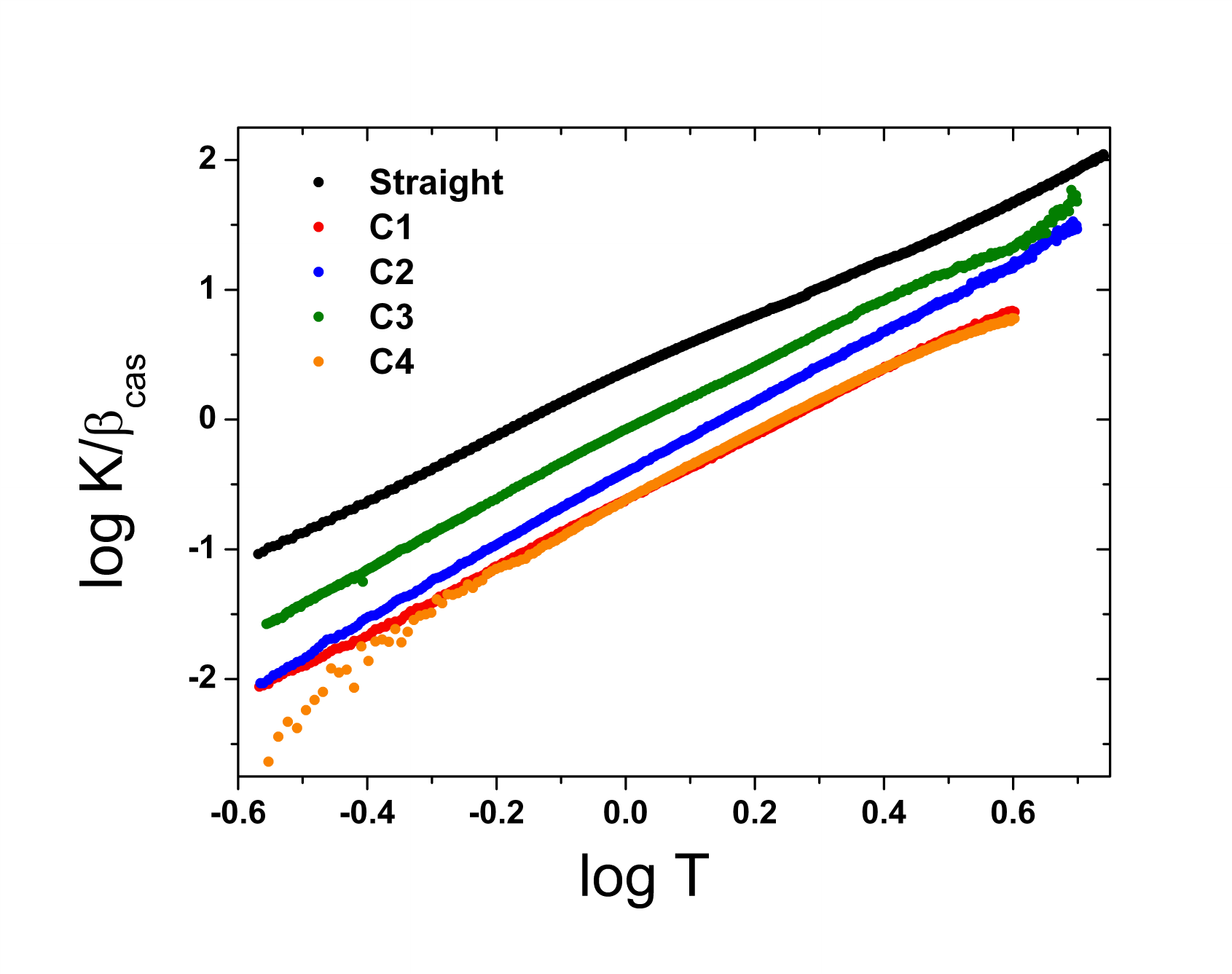}
 \end{center}
 \caption{(colour online) Log-log plot of the normalized thermal conductance of the straight and four corrugated nanowires versus the temperature. The normalization coefficient $\beta_{Cas}$ corresponds to the Casimir limit value of the thermal conductance at 1~K.} 
 \label{fig3}
\end{figure}

where $p$ is a parameter describing the probability for a phonon to be specularly reflected when hitting the surfaces. The value of this parameter will depend on both the wavelength of the phonons and the roughness of the wires.
In this model, the phonon boundary scattering ranges between the purely diffusive case when $p=0$ (i.e., the Casimir limit) and the purely specular one when $p=1$. The experimental thermal conductance can then be expressed as a function of the effective MFP $\Lambda_{eff}$:
\begin{equation}
K=K_{Cas}\Lambda_{eff}/\Lambda_{Cas}.
\label{keff}
\end{equation}
We have normalized the experimental data with respect to the Casimir coefficient $\beta_{Cas}$ (according to equation~\ref{Caseq} and equation~\ref{zimfp}) in order to extract the $\Lambda_{eff}$ MFP and deduce the corresponding value of the parameter $p$ for each corrugated nanowire. Fig.~\ref{fig3} reports the thermal conductance plotted as $K/\beta_{Cas}$ versus temperature. This normalization allows a direct comparison between the thermal transports of each nanowire. From this plot, it is already patent that the presence of corrugation strongly reduces the thermal transport.

To go further in the data analysis, a fitting of the data $K/\beta_{Cas}$ versus temperature has been done. From equations \ref{Caseq} and \ref{zimfp}, the normalized conductance can be expressed as
\begin{equation}
log (K/\beta_{Cas}) = a+b\hspace{0.1cm}log(T),
\label{kkcas1}
\end{equation}
where 
\begin{equation}
(K/\beta_{Cas})=\frac{1+p}{1-p} T^{b}.
\label{kkcas2}
\end{equation}
From these two equations, the power law factor $b$ (close to 3 for 3D materials), the value of the MFP $\Lambda_{eff}$, and the parameter $p$ have been extracted; the data are collected in the Table~\ref{tablo}. 
 
\begin{figure}
\begin{center}
 \includegraphics[width=8.5cm]{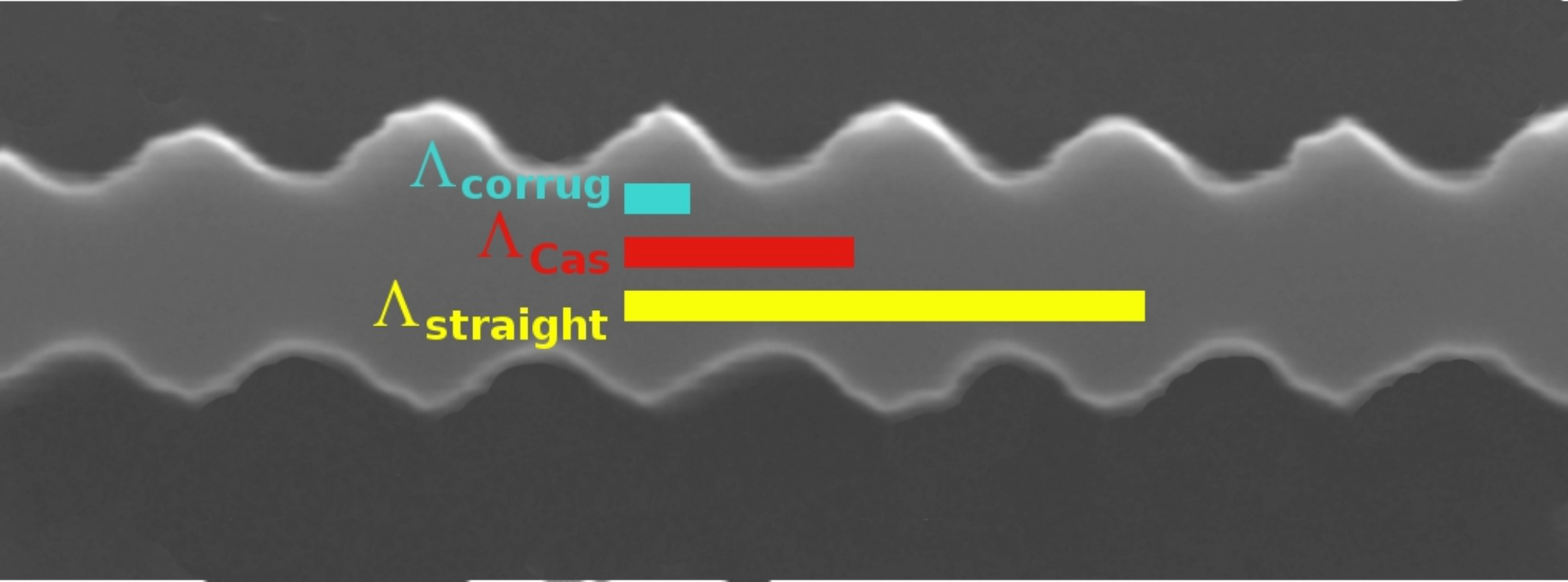}
 \end{center}
 \caption{(colour online) The Casimir MFP ($\Lambda_{Cas}=220$nm), the effective MFP in the straight nanowire ($\Lambda_{straight}=454$nm) and the corrugated nanowire C1 ($\Lambda_{corrug}=57$nm) are compared to the actual geometry of the corrugated nanowire.} 
 \label{fig4}
\end{figure}

As it can be seen in this table, the MFPs of the corrugated nanowires are strongly reduced as compared to the one of the straight nanowire. The reduction by a factor of almost ten is drastic, and can only be attributed to the presence of the corrugation. Moreover, for samples C1, C2 and C4, the effective MFP is patently smaller than the smallest section of the nanowires, demonstrating a phonon flux weaker than the one expected in the Casimir limit \cite{carrete}. Fig.~\ref{fig4} illustrates the different MFPs ($\Lambda_{straight}$, $\Lambda_{corrug}$ and $\Lambda_{Cas}$) at the scale of the corrugated nanowires. As demonstrated in reference \cite{PRBHeron,Heron1} large MFPs in the straight geometry are a signature of the presence of ballistic phonons \cite{Heron2}: those may be particularly affected by the corrugation. Finally, the significant reduction in the MFP is also illustrated by the $p$ value, which is either small or negative. As proposed by Moore \textit{et al.} in \cite{sawtooth}, this reduction may be the signature of the presence of strong phonon trapping in the wire due to multiple phonon scattering or even backscattering. However, the negative parameter $p$ obtained for samples C1, C2 and C4 does not seem to be physically sound.

In the Ziman model, the paramater $p$ is the probability for one phonon reflection to be diffusive or specular. It does not take into account the direction of the reflected phonons. In our case, as in some recent experiments and theorical studies \cite{carrete,sawtooth,hochbaum,shi2003}, multiple reflections and especially backscattering may have a significant impact on the phonon transport.

If, as suspected, the reduction in the MFP is due to the presence of multiple phonon scattering and backscattering, this should turn up in Monte Carlo simulations. Indeed, multiple scattering effects do not involve the wave aspect of the phonon but only its corpuscular behaviour, something which is particularly well handled by a Monte Carlo analysis, which has been carried out here according to ray tracing principles \cite{rajabpour}. 

\begin{figure}
\begin{center}
 \includegraphics[width=8.5cm]{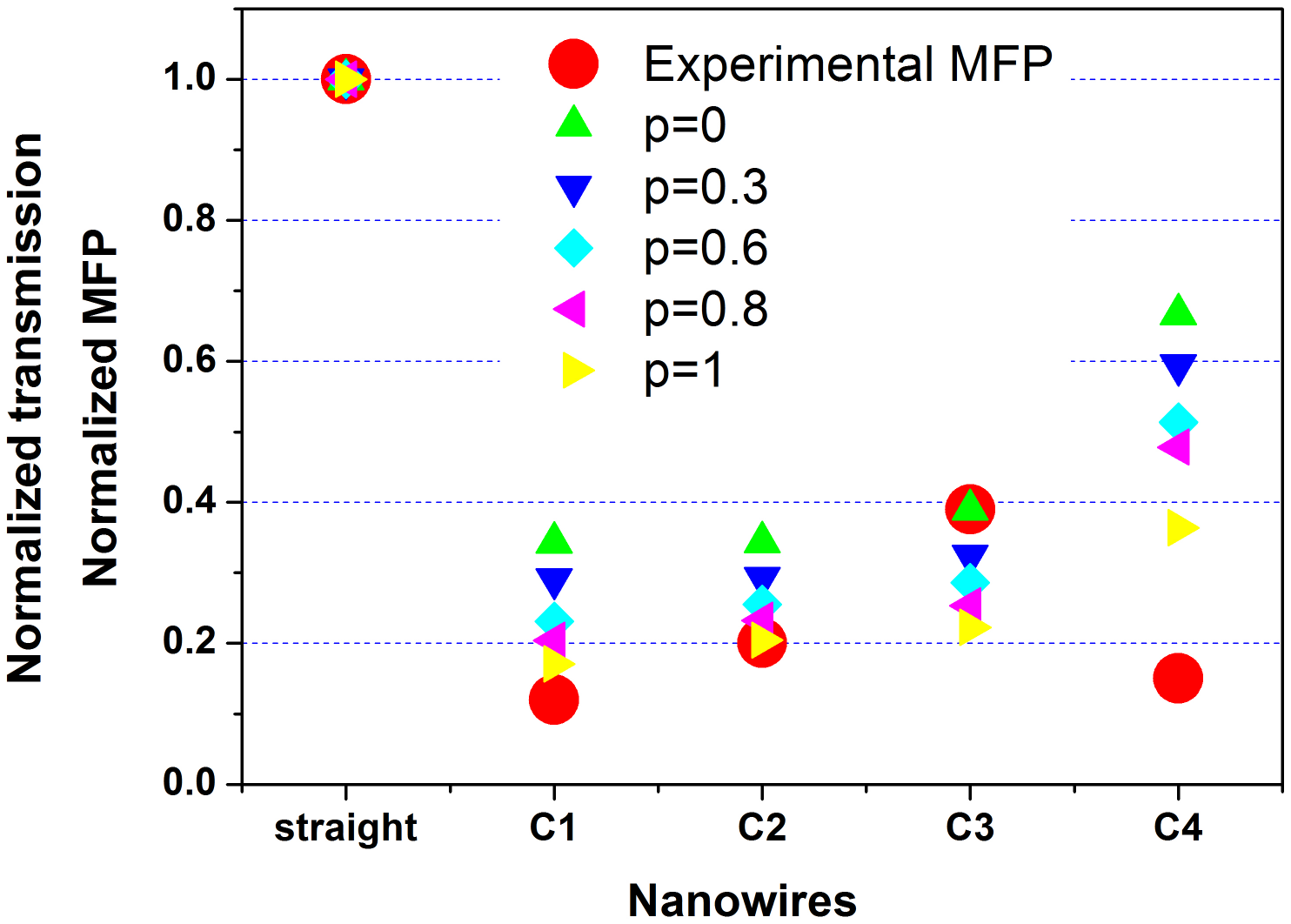}
 \end{center}
 \caption{(colour online) The red full circles are the experimental MFP normalized by the MFP in the straight nanowire. The Monte Carlo simulations have been performed with different parameters $p$ (see the figure legend). The transmissions obtained by the MC simulations are normalized to the transmission in the straight wire.} 
 \label{MC}
\end{figure}

In the grey approximation, when the frequency dependence is discarded, phonons are emitted from one side of the wire with a random initial direction and a random initial position. The phonons are then tracked until leaving the wire through one of its sides. The ratio between the number of phonons reaching the other side $N_{through}$ to the initial number of phonons $N$ defines the transmission $T=N_{through}/N$. The only scattering mechanism is generated by the surfaces and was modeled by different specularity parameters ranging from $p=0$ to $p=1$ as reported in Fig.~\ref{MC}; the value of the parameter $p$ derived from the thermal measurement of the straight nanowire is $p=0.6$. The transmission coefficients can then be related to the MFP through the relation \cite{bera,Lund}

\begin{equation}
T=\frac{1}{1+\frac{3}{4}\frac{L}{\Lambda}}.
\label{trans}
\end{equation}

In this equation, $\Lambda$ stands for any MFP and $L$ is the length of the wire.
Following equation~\ref{trans}, a direct correspondence is obtained between the transmission coefficient calculated from the Monte Carlo simulations and the MFP. The simulations reveal a significant decrease in the phonon transmission for the corrugated nanowires, corresponding to the backscattering of phonons towards the entrance cross section. The comparison between the calculated transmissions and experimental MFPs are presented in Fig.~\ref{MC}. The simulations provide the correct trend when varying the roughness of the nanowires through the parameter $p$: for instance, $p$ is close to 1 for samples C1, C2 and C4 and a parameter $p$ close to 0 has to be chosen for sample C3. As a general conclusion, the MC simulations clearly support the experimental fact that in the presence of corrugation, the MFP is indeed strongly reduced. This reduction may be attributed to the occurence of phonon multiple scattering, which is inherent to the ray tracing method.

Sample C3 presents a smaller mean free path reduction: the MFP and the parameter $p$ are smaller than that of the straight nanowire but not by as much as in the other samples (C1, C2 and C4). Less  multiple scattering can be evidenced in this nanowire. This may be due to the nanowire surfaces, which can be rougher than in their counterparts, restoring diffusive transport closer to the Casimir limit, as confirmed by the MC simulations. The behavior of the sample C4 cannot be described by the simulations; indeed, the Monte Carlo predictions are not able to capture the range of the experimental data. However, other mechanisms may be responsible for such an MFP reduction. Two scenarios of phonon scattering have recently been proposed which could explain a thermal transport far below the Casimir limit: coherent effects, involving correlated multiple scatterings of phonons \cite{sinha}, or strong phonon scattering induced by highly perturbated surfaces (core defects and surface ripples) \cite{he}; validating such scenarios would require further experimental investigations.

To conclude, the major result of this research resides in the significant reduction in the phonon MFP in monocrystalline silicon nanowires related to the presence of multiple reflections induced by corrugated surfaces. As a consequence, heat transport has also strongly decreased below the Casimir limit. The hypothesis of multiple reflections for phonons is largely confirmed by the ray tracing analysis showing a strong diminution in the phonon transmission. In the best case, the effective MFP has been reduced by more than a factor of nine from that of the straight nanowire MFP. This is a clear signature that phonons are being backscattered due to the presence of the corrugated surfaces. This drastic MFP reduction is the demonstration that a monocrystalline material can be turned into a phonon glass-like material merely by introducing corrugated surfaces. Corrugations even allow breaking the Casimir limit of thermal phonon transport in low dimensional systems. The thermal properties of these metamaterials is of particular interest in the quest for decreasing phonon thermal transport without affecting the electronic properties, which significantly improves the performance of thermoelectrics.


\begin{thebibliography}{99}
\bibitem{note3}
When using the 3 omega technique for the measurement of thermal conductance of suspended nanowire, only half the length has to be considered. The hottest point is indeed located in the middle of suspended system.
\bibitem{note2}
The sensitivity is on the order of a femtoWatt/K ($10^{-15}$~W/K) for a noise equivalent power of $10^{-14}$~W/$\sqrt{\rm{Hz}}$. With temperature oscillations of 10~mK, this experiment is able to quantify the attoJoule ($10^{-18}$~J) at 5~Hz and at the lowest temperature.
\end{thebibliography}
\end{document}